\documentclass[12pt]{article}

\usepackage{amsfonts,amsmath,amssymb}

\usepackage{tikz}

\newcommand{\ra}{\rightarrow}
\newcommand{\Hom}{{\rm Hom}}

\newcommand{\CC}{{\mathbb C}}
\newcommand{\ZZ}{{\mathbb Z}}
\newcommand{\RR}{{\mathbb R}}

\newcommand{\PP}{{\mathbb P}}

\newcommand{\Tr}{{\rm Tr}}

\newcommand{\sign}{{\it sign}}

\begin{document}

\title{Fermionic Symmetry Protected Topological Phases and Cobordisms}

\author{Anton Kapustin\footnote{California Institute of Technology, Pasadena, CA}
\\ Ryan Thorngren\footnote{University of California, Berkeley, CA}
\\ Alex Turzillo\footnote{California Institute of Technology, Pasadena, CA}
\\ Zitao Wang\footnote{California Institute of Technology, Pasadena, CA}}

\maketitle

\abstract{It has been proposed recently that interacting Symmetry Protected Topological Phases can be classified using cobordism theory. We test this proposal in the case of Fermionic SPT phases with $\ZZ_2$ symmetry, where $\ZZ_2$ is either time-reversal or an internal symmetry. We find that cobordism classification correctly describes all known Fermionic SPT phases in space dimension $D\leq 3$ and also predicts that all such phases can be realized by free fermions. In higher dimensions we predict the existence of inherently interacting fermionic SPT phases.}

\section{Introduction}

Classification of Symmetry Protected Topological Phases has been a subject of intensive activity over the last few years. In the case of free fermions, a complete  classification has been achieved in \cite{Furusakietal, Kitaev} using such ideas as Anderson localization and K-theory. In the case of bosonic systems, all SPT phases are intrinsically interacting, so one has to use entirely different methods. Interactions are also known to affect fermionic SPT phases \cite{Fid10, CFV, Wang14, LevinGu}.  Recently it has been proposed that cobordism theory can provide a complete classification of both bosonic and fermionic interacting SPT phases in all dimensions. This improves on the previous proposal that group cohomology classifies interacting bosonic SPT phases \cite{groupcohomology}, while ``group supercohomology'' \cite{GuWen} classifies interacting fermionic SPT phases. For bosonic systems with time-reversal and $U(1)$ symmetries the cobordism proposal has been tested in \cite{Kapcob1} and \cite{Kapcob2} respectively. Cobordism theory has been found to describe all known bosonic SPT phases with such symmetries in $D\leq 3$. In this paper we test the proposal further by studying fermionic SPT phases with $\ZZ_2$ symmetry.

The $\ZZ_2$ symmetry in question can be either unitary or anti-unitary. In the former case we will assume that the symmetry is internal (does not act on space-time). In the latter case it must reverse the direction of time, so we will call it time-reversal symmetry. In either case, the generator can square either to $1$ or to $(-1)^F$ (fermion parity). Fermionic SPT phases with time-reversal symmetry are also known as topological superconductors, so in particular we describe a classification scheme for interacting topological superconductors.

Compared to the bosonic case, fermionic SPT phases present several related difficulties. First of all, one needs to decide what one means by a fermionic system. In a continuum Lorentz-invariant field theory, anti-commuting fields are also spinors with respect to the Lorentz group, but condensed matter systems are usually defined on a lattice and lack Lorentz invariance on the microscopic level. Thus the connection between spin and statistics need not hold. A related issue is that all fermionic systems have $\ZZ_2$ symmetry called fermionic parity, usually denoted $(-1)^F$. But all observables, including the Hamiltonian and the action, are bosonic, i.e. invariant under $(-1)^F$. In a sense, every fermionic system has a $\ZZ_2$ gauge symmetry, which means that the partition function must depend on a choice of a background $\ZZ_2$ gauge field. It is tempting to identify this gauge field with the spin structure. However, it is not clear how a spin structure should be defined for a lattice system, except in the case of toroidal geometry.\footnote{In 2d, there is a good combinatorial description of spin structures via so called Kasteleyn orientations \cite{CR}. But a generalization of this construction to higher dimensions is unknown.}

Instead of dealing with all these difficult questions, in this paper we take a more ``phenomenological'' approach: we make a few assumptions about the long-distance behavior of SPT phases which parallel those for bosonic SPT phases, and then test these assumptions by comparing the results in space-time dimensions $d\leq 4$ with those available in the condensed matter literature. For various reasons, we limit our selves to the cases of no symmetry, time-reversal symmetry, and unitary $\ZZ_2$ symmetry. Having found agreement with the known results, we make a conjecture about the classification of fermionic SPT phases with any symmetry group $G$. 

R. T. would like to thank Xie Chen for patiently answering his questions and Rob Kirby for an enlightening discussion. A. K. would like to acknowledge conversations with Alexei Kitaev and Zhengcheng Gu. The work of A. K., A. T., and Z. W.  was supported in part DOE grant  DE-FG02-92ER40701

\section{Spin and Pin structures}

A smooth oriented $d$-manifold $M$ equipped with a Riemannian metric is said to have a spin structure if the transition functions for the tangent bundle, which take values in $SO(d)$, can be lifted to $Spin(d)$ while preserving the cocycle condition on triple overlaps of coordinate charts. Let us unpack this definition. On a general manifold one cannot choose a global coordinate system, so one covers $M$ with coordinate charts $U_i,$ $i\in I$. If over every coordinate chart $U_i$ one picks an orthonormal basis of vector fields with the correct orientation, then on double overlaps $U_{ij}=U_i\bigcap U_j$ they are related by transition functions $g_{ij}$ which take values in the group $SO(d)$ and satisfy on $U_{ijk}=U_i\bigcap U_j \bigcap U_k$ the cocycle condition:
\begin{equation}\label{gcocycle}
g_{ij}g_{jk}=g_{ik}.
\end{equation}
The group $SO(d)$ has a double cover $Spin(d)$, i.e. one has $SO(d)=Spin(d)/\ZZ_2$. One can lift every smooth function $g_{ij}:U_{ij}\ra SO(d)$ to a smooth function $h_{ij}:U_{ij}\ra Spin(d)$, with a sign ambiguity. Thus on every $U_{ijk}$ one has
\begin{equation}\label{hcocycle}
h_{ij} h_{jk}=\pm h_{ik}.
\end{equation}
$M$ has a spin structure if and only if one can choose the functions $h_{ij}$ so that the sign on the right-hand side is $+1$ for all $U_{ijk}$. We also identify spin structures which are related by $Spin(d)$ gauge transformations:
$$
h_{ij}\mapsto h'_{ij}=h_i h_{ij} h_j^{-1}, \quad h_i:U_i\ra Spin(d).
$$
A spin structure allows one to define Weyl spinors on $M$.

For $d<4$ every oriented $d$-manifold admits a spin structure, but it is not unique, in general. Namely, given any spin structure, one can modify it by multiplying every $h_{ij}$ by constants $\zeta_{ij}=\pm 1$ satisfying
$$
\zeta_{ij}\zeta_{jk}=\zeta_{ik}.
$$
Such constants define a Cech 1-cochain on $M$ with values in $\ZZ_2$. The same data also parameterize $\ZZ_2$ gauge fields on $M$, thus any two spin structures differ by a $\ZZ_2$ gauge field. It is easy to see that gauge fields differing by $\ZZ_2$ gauge transformations lead to equivalent transformations of spin structures, so the number of inequivalent spin structures is equal to the order of the Cech cohomology group $H^1(M,\ZZ_2)$, whose elements label gauge-equivalence classes of $\ZZ_2$ gauge fields.

In dimension $d>3$ not every oriented manifold admits a spin structure. For example, the complex projective plane $\CC\PP^2$ does not admit a spin structure. Nevertheless, if a spin structure on $M$ exists, the above argument still shows that the number of inequivalent spin structures is given by $|H^1(M,\ZZ_2)|$. The necessary and sufficient condition for the existence of a spin structure is the vanishing of the 2nd Stiefel-Whitney class $w_2(M)\in H^2(X,\ZZ_2)$. This condition is purely topological and thus does not depend on the choice of Riemannian metric on $M$. 

If $M$ is not oriented, the transition functions $g_{ij}$ take values in $O(d)$ rather than $SO(d)$. They still satisfy (\ref{gcocycle}). An analog of  $Spin$ group in this case is called a $Pin$ group. In the absence of orientation, fermions transform in a representation of the $Pin$ group. In fact, for all $d>0$ there exist two versions of the $Pin$ group called $Pin^+(d)$ and $Pin^-(d)$. They both have the property $Pin^\pm(d)/\ZZ_2=O(d)$. The difference between $Pin^+$ and $Pin^-$ is the way a reflection of any one of coordinate axis is realized on fermions. Let $r\in O(d)$ be such a reflection. It satisfies $r^2=1$. If $\tilde r\in Pin^\pm(d)$ is a pre-image of $r$, it can satisfy either $\tilde r^2=1$ or $\tilde r^2=-1$. The first possibility corresponds to $Pin^+$, while the second one corresponds to $Pin^-$. 

If we are given an unoriented $d$-manifold $M$, we can ask whether it admits $Pin^+$ or $Pin^-$ structures (that is, lifts of transition functions to either $Pin^+(d)$ or $Pin^-(d)$ so that the condition (\ref{hcocycle}) on triple overlaps is satisfied). The conditions for this are again topological: in the case of $Pin^+$ it is the vanishing of $w_2(M)$, while in the case of $Pin^-$ it is the vanishing of $w_2(M)+w_1(M)^2$. Note that if $M$ happens to be orientable, then $w_1(M)=0$, so the two conditions coincide and reduce to the condition that $M$ admit a $Spin$ structure. 

Note that these topological conditions are nontrivial already for $d=2$. More precisely, for $d=2$ one has a relation between Stiefel-Whitney classes $w_1^2+w_2=0$, so every 2d manifold admits a $Pin^-$ structure, but not necessarily a $Pin^+$ structure. For example the real projective plane $\RR\PP^2$ admits only $Pin^-$ structures, while the Klein bottle admits both $Pin^+$ and $Pin^-$ structures. Similarly, not every 3-manifold admits a $Pin^+$ structure, but all 3-manifolds admit a $Pin^-$ structure. 

\section{Working assumptions}

We assume that fermionic SPTs in $d$ space-time dimensions without time-reversal symmetry can be defined on any oriented smooth $d$-manifold $M$ equipped with a spin structure. Similarly, we assume that fermionic SPTs with time-reversal symmetry can be defined on any smooth manifold $M$ equipped with a $Pin^+$ or $Pin^-$ structure (we will see below that $Pin^+$ corresponds to $T^2=(-1)^F$ while $Pin^-$ corresponds to $T^2=1$). If there are additional symmetries beyond $(-1)^F$ and time-reversal, $M$ can carry a background gauge field for this symmetry. 

We also assume that given such $M$, a long-distance effective action is defined. The action is related to the partition function by $Z=\exp(2\pi i S_{eff})$, thus $S_{eff}$ is defined modulo integers. The trivial SPT phase corresponds to the trivial (zero) action. The effective action is additive under the disjoint union of manifolds. It also changes sign under orientation-reversal. In the case of SPT phases with time-reversal symmetry, this implies  $2S_{eff}\in\ZZ$. 

The effective action, in general, is not completely topological: it may depend on the Levi-Civita connection on $M$. Such actions are gravitational Chern-Simons terms and can exist if $d=4k-1$. Since we will be interested only in low-dimensional SPT phases, the only case of interest is $d=3$. The correspond gravitational Chern-Simons term has the form
$$
S_{CS}=\frac{k}{192\pi}\int \Tr(\omega d\omega+\frac{2}{3}\omega^3),
$$
where the trace is in the adjoint representation of $SO(3)$. Note that such a term makes sense only on an orientable 3-manifold and therefore can appear only if the symmetry group of the SPT phase does not involve time reversal. 

In the bosonic case, one can show that $k$ must be an integral multiple of $16$. In the fermionic case, $k$ can be an arbitrary integer. The quantization of $k$ is explained in the appendix. 

The physical meaning of $S_{CS}$ is that it controls the thermal Hall response of the SPT phases \cite{RG}. The thermal Hall conductivity is  proportional to $k$ \cite{RG}:
$$
\kappa_{xy}=\frac{k\pi k_B^2 T}{12\hbar},
$$ 
where $T$ is the temperature and $k_B$ is the Boltzmann constant.
Thus for both bosonic and fermionic SPT phases the quantity $\kappa_{xy}/T$ is quantized, but in the fermionic case the quantum is smaller than in the bosonic case by a factor $16$. This is derived in the appendix.

SPT phases with a particular symmetry form an abelian group, where the group operation amounts to forming the composite system. The effective action is additive under this operation. Taking the inverse corresponds to applying time-reversal to the SPT phase. The effective action changes sign under this operation. Thus the effective action can be regarded as a homomorphisms from the set of SPT phases to $\RR/\ZZ\simeq U(1)
$. 

The difference of two SPT phases with the same thermal Hall conductivity is an SPT phase with zero thermal Hall conductivity. Thus it is sufficient to classify SPT phases with zero thermal Hall conductivity. In such a case the action is purely topological. Our final assumption is that this topological action depends only on the bordism class of $M$. Equivalently, we assume that if $M$ is a boundary of some $d+1$-manifold with the same structure ($Spin$ or $Pin^\pm$, as the case may be), then $S_{eff}$ vanishes. This assumption is supposed to encode locality.

\section{Fermionic SPT phases without any symmetry}

\begin{table} 
\caption{$Spin$ and $Pin^\pm$ Bordism Groups} 
\centering
\renewcommand{\arraystretch}{1.1}
\begin{tabular}{c | c c c c}
\\ [-2.5ex] \hline\hline \\ [-2.5ex]
$d=D+1$ & $\Omega_{d}^{Spin}(pt)$ & $\Omega_d^{Pin^-}(pt)$ & $\Omega_d^{Pin^+}(pt)$ & $\Omega_d^{Spin}(B\ZZ_2)$ \\ [0.5 ex]
\hline \\ [-2.5ex]
1 & $\ZZ_2$ & $\ZZ_2$ & 0 & $\ZZ_2^2$ \\ 
2 & $\ZZ_2$ & $\ZZ_8$ & $\ZZ_2$ & $\ZZ_2^2$ \\ 
3 & 0 & 0 & $\ZZ_2$ & $\ZZ_8$\\ 
4 & $\ZZ$ & 0 & $\ZZ_{16}$ & $\ZZ$\\
5 & 0 & 0 & 0 & 0\\
6 & 0 & $\ZZ_{16}$ & 0 & 0\\
7 & 0 & 0 & 0 & $\ZZ_{16}$\\
8 & $\ZZ^2$ & $\ZZ_2^2$ & $\ZZ_2\times\ZZ_{32}$ & $\ZZ^2$\\
9 & $\ZZ_2^2$ & $\ZZ_2^2$ & $ 0 $ & $\ZZ_2^4$ \\
10 & $\ZZ_2^2 \times \ZZ$ & $\ZZ_{2}\times\ZZ_{8}\times\ZZ_{128}$ & $\ZZ_2^3$ & $\ZZ_2^4 \times \ZZ$\\
\hline  
\end{tabular} 
\label{table:cobord} 
\end{table}

\begin{table} 
\caption{Interacting Fermionic SPT Phases} 
\centering
\renewcommand{\arraystretch}{1.1}
\begin{tabular}{c | c c c c}
\\ [-2.5ex] \hline\hline \\ [-2.5ex]
$d=D+1$ & no symmetry & $T^2 = 1$ & $T^2 = (-1)^F$ & unitary $\ZZ_2$ \\ [0.5ex]
\hline \\ [-2.5ex]
1 & $\ZZ_2$ & $\ZZ_2$ & 0 & $\ZZ_2^2$ \\ 
2 & $\ZZ_2$ & $\ZZ_8$ & $\ZZ_2$ & $\ZZ_2^2$ \\ 
3 & $\ZZ$ & 0 & $\ZZ_2$ & $\ZZ_8\times\ZZ$\\ 
4 & 0 & 0 & $\ZZ_{16}$ & 0\\
5 & 0 & 0 & 0 & 0\\
6 & 0 & $\ZZ_{16}$ & 0 & 0\\
7 & $\ZZ^2$ & 0 & 0 & $\ZZ_{16}\times\ZZ^2$\\
8 & 0 & $\ZZ_2^2$ & $\ZZ_2\times\ZZ_{32}$ & 0\\
9 & $\ZZ_2^2$ & $\ZZ_2^2$ & $ 0 $ & $\ZZ_2^4$ \\
10 & $\ZZ_2^2$ & $\ZZ_{2}\times\ZZ_{8}\times\ZZ_{128}$ & $\ZZ_2^3$ & $\ZZ_2^4$\\
\hline  
\end{tabular} 
\label{table:nonlin} 
\end{table}

\begin{table} 
\caption{Free Fermionic SPT Phases} 
\centering
\renewcommand{\arraystretch}{1.1}
\begin{tabular}{c | c c c}
\\ [-2.5ex] \hline\hline \\ [-2.5ex]
$d=D+1 \ {\rm mod}\ 8$ & no symmetry & $T^2 = 1$ & $T^2 = (-1)^F$ \\ [0.5ex]
\hline \\ [-2.5ex]
1 & $\ZZ_2$ & $\ZZ_2$ & $0$ \\ 
2 & $\ZZ_2$ & $\ZZ$ & $\ZZ_2$ \\ 
3 & $\ZZ$ & $0$ & $\ZZ_2$ \\ 
4 & $0$ & $0$   & $\ZZ$ \\
5 & $0$ & $0$   & $0$ \\
6 & $0$ & $\ZZ$ & $0$ \\
7 & $\ZZ$ & $0$ & $0$  \\
8 & $0$ & $\ZZ_2$ & $\ZZ$ \\
\hline  
\end{tabular} 
\label{table:free} 
\caption{Classification of free fermionic SPT phases according to \cite{Furusakietal} and \cite{Kitaev}. The ``no symmetry'' case corresponds to class D, the case $T^2=1$ corresponds to class BDI, the case $T^2=(-1)^F$ corresponds to class DIII.}
\end{table}

In the case when the only symmetry is $(-1)^F$, the manifold $M$ can be assumed to be a compact oriented manifold with a spin structure. As explained above, without loss of generality we may assume that the action is purely topological (depends only on the spin bordism class of $M$). Thus possible effective actions in space-time dimension $d$ are classified by elements of the group $\Hom(\Omega_d^{Spin}(pt),U(1))$, where $\Omega_d^{Spin}(pt)$ is the group of bordism classes of spin manifold of dimension $d$. 

The spin bordism groups $\Omega_d^{Spin}(pt)$ have been computed by Anderson, Brown, and Peterson \cite{Anderson67}. In low dimensions, one gets
\begin{equation}
\Omega_{1}^{Spin}(pt) = \mathbb{Z}_2, \ \ \ \Omega_{2}^{Spin}(pt) = \mathbb{Z}_2, \ \ \ \Omega_{3}^{Spin}(pt) = 0, \ \ \ \Omega_{4}^{Spin}(pt) = \mathbb{Z}, \nonumber \\ 
\end{equation}
If a bordism group contains a free part, its Pontryagin dual has a $U(1)$ factor. This means that the corresponding effective action can depend on a continuous parameter. If we want to classify SPT phases up to homotopy, we can ignore such parameters. This is equivalent to only considering the torsion subgroup of $\Omega_d^{Spin}(pt)$. Thus we propose that SPT phases in dimension $d$  are classified by elements of the Pontryagin dual of the torsion subgroup of $\Omega_d^{Spin}(pt)$. We will denote this group $\Omega^{d,tors}_{Spin}(pt)$.

The groups $\Omega_d^{Spin}$ are displayed in Table 1. The classification of interacting fermionic SPT phases can be deduced from it in the manner just described and is displayed in Table 2. For comparison, the classification of free fermionic SPT phases described in \cite{Furusakietal} and \cite{Kitaev} is shown in Table 3. We see that there are nontrivial interacting fermionic SPT phases with zero thermal Hall response in $D=0\text{ and }1$ but not in $D=2\text{ and }3$. However, for $D=2$ there is a phase with a nontrivial thermal Hall response; it is also present in the table of free fermionic SPT phases. In higher dimensions the number of phases grows rapidly. For instance, the effective action can be any combination of the Stiefel-Whitney numbers modulo $w_1$ and $w_2$ (such effective actions correspond to fermionic phases which are independent of the spin structure on $M$ and thus can also be regarded as bosonic phases).

Let us consider the cases $d=1$ and $d=2$ in slightly more detail. For $d=1$, there is only one connected closed manifold, namely, the circle. There are two spin structures on a circle: the periodic one and the anti-periodic one. The nontrivial effective action assigns a different sign to each spin structure and is multiplicative over disjoint unions. From the point of view of quantum mechanics, such an effective action corresponds to the $d=1$ SPT phase whose unique ground state is fermionic. 

In two space-time dimensions, the situation is more complicated. Spin structures on an oriented 2d manifold $X$ can be thought of as $\ZZ_2$ valued quadratic forms on $H_1(X,\ZZ_2)$ satisfying $q(x+y) = q(x) + x\cap y + q(y) \mod 2$, where $x\cap y$ denotes the $\ZZ_2$ intersection pairing. The bordism invariant is the Arf invariant, which is the obstruction to finding a Lagrangian subspace for this quadratic form. The effective action for the nontrivial SPT phase in $D=1$ is given by the Arf invariant \cite{KirbyTaylor}
\begin{equation}\label{ABK}
S(q) = \frac{1}{\sqrt{|H^1(X,\ZZ_2)|}} \sum_{A \in H^1(X,\ZZ_2)} \exp(2\pi i q(A)/2).
\end{equation}
Another way to describe the Arf invariant is to consider zero modes for the chiral Dirac operator. Their number modulo 2 is an invariant of the spin structure and coincides with the Arf invariant \cite{stringlit}. In string theory, spin structures for which the Arf invariant is even (resp. odd) are called even (resp. odd). 

The spin cobordism classification is consistent with existing results in condensed matter literature. Fidkowski and Kitaev \cite{Fid10} have considered the Majorana chain with just fermion parity. There are two distinct phases: one where all sites are decoupled and unoccupied in the unique ground state and one with dangling Majorana operators which can be paired into a gapless Dirac mode representing a two-fold ground state degeneracy. In the absense of any symmetry beyond $(-1)^F$, a four-fermion interaction can gap out the dangling modes in pairs, so these are the only two phases. 

\section{Fermionic SPT phases with time-reversal symmetry}

\subsection{General considerations}

In the presence of time-reversal symmetry, the manifold $M$ can be unorientable. As discussed in section 2, there are two distinct unoriented analogs of a spin structure, called $Pin^+$ and $Pin^-$ structures. They should correspond to the two possibilities for the action of time-reversal: $T^2=1$ and $T^2=(-1)^F$. 

Naively, it seems that $T^2=1$ should correspond to $Pin^+$ and $T^2=(-1)^F$ should correspond to $Pin^-$. Indeed, for $Pin^+$ the reflection of a coordinate axis acts on a fermion by an element $\tilde r$ satisfying $\tilde r^2=1$, while for $Pin^-$ it acts by $\tilde r$ satisfying $\tilde r^2=-1$. However, one should take into account that the groups $Pin^\pm$ are suitable for space-time of Euclidean signature. A reflection of a coordinate axis in Euclidean space is related to time-reversal by a Wick rotation. Let $r$ be a reflection of the coordinate axis which is to be Wick-rotated. The corresponding element of $Pin^\pm$ acts on the fermions by a Dirac matrix $\gamma_d$ which satisfies $\gamma_d^2=\pm 1$. Wick rotation amounts to $\gamma_d\mapsto i\gamma_d$, hence $Pin^+$ corresponds to $T^2=(-1)^F$, while $Pin^-$ corresponds to $T^2=1$. This identification will be confirmed by the comparison with the results from the condensed matter literature.

\subsection{$T^2=(-1)^F$}

We propose that interacting fermionic SPT phases protected by time-reversal symmetry $T$ with $T^2 = (-1)^F$ are classified by elements of 
$$\Omega^d_{Pin^+}(pt)=\Hom(\Omega_d^{Pin^+}(pt),U(1)).$$
We will call this group the $Pin^+$ cobordism group with $U(1)$ coefficients.

The $Pin^+$ bordism groups have been computed by Kirby and Taylor \cite{Kirby90}
\begin{equation}
\Omega_{1}^{Pin^+}(pt) = 0, \ \ \ \Omega_{2}^{Pin^+}(pt) = \mathbb{Z}_2, \ \ \ \Omega_{3}^{Pin^+}(pt) = \ZZ_2, \ \ \ \Omega_{4}^{Pin^+}(pt) = \mathbb{Z}_{16}, \nonumber \\ 
\end{equation}
$Pin^+$ bordism groups grow quickly with dimension, soon having multiple cyclic factors.

In one space-time dimension, the $Pin^+$ cobordism group vanishes. This is easily interpreted in physical terms. Recall that without time-reversal symmetry, the ground state can be bosonic or fermionic, and the latter possibility corresponds to a nontrivial fermionic $d=1$ SPT phases. However, if time-reversal symmetry $T$ with $T^2=(-1)^F$ is present, fermionic states are doubly-degenerate, and since by definition the ground state of an SPT phase are non-degenerate, the ground state cannot be fermionic. 

In two space-time dimensions, there is an isomorphism
$$
\Omega_2^{Pin^+}(pt) \to \Omega_2^{Spin}(pt),
$$
see \cite{KirbyTaylor}. The isomorphism arises from the fact that a $Pin^+$ structure on an unoriented manifold induces a spin structure on its orientation double cover.   Thus there is a unique nontrivial fermionic SPT phase in $d=2$, and the corresponding effective action is simply the action \eqref{ABK} on the orientation double cover:
$$
S(q) = \frac{1}{\sqrt{|H^1(\tilde X,\ZZ_2)|}} \sum_{A \in H^1(\tilde X,\ZZ_2)} e^{2\pi i q(A)/2}.
$$
The classification of the free fermionic SPTs in $d=2$ also predicts a unique nontrivial phase with time-reversal symmetry $T^2=(-1)^F$ \cite{Furusakietal,Kitaev}. It can be realized by a time-reversal-invariant version of the Majorana chain and is characterized by the presence of a pair of dangling Majorana zero modes on the edge.

In three space-time dimensions, a similar map is not an isomorphism, as $\Omega_3^{Spin} = 0$. However, there is a map
\begin{equation}\label{wonecup}
[\cap w_1]:\Omega_3^{Pin^+} \to \Omega_{2}^{Spin}
\end{equation}
taking a $Pin^+$ manifold to a codimension 1 submanifold Poincar\'e dual to the orientation class $w_1$. This submanifold is defined to be minimal for the property that the complement can be consistently oriented. With this choice of partial orientation, crossing this submanifold reverses the orientation, so it can be thought of as a time-reversal domain wall. For $Pin^+$ 3-manifolds, we have $w_1^2 = 0$, so this domain wall is oriented and inherits a Spin structure from the ambient spacetime.

The map (\ref{wonecup}) is an isomorphism \cite{KirbyTaylor}. From the physical viewpoint 
this means that away from the time-reversal domain walls the SPT is trivial and the boundary can be gapped, but on the domain walls there is a $d=2$  fermionic SPT, the Majorana chain, so at locations where the domain walls meet the boundary there are Majorana zero modes. This is a special case of a construction of SPT phases discussed in the bosonic case in \cite{ChenLuVishwanath}. One starts with a system with symmetry $G$ in a trivial phase, breaks the $G$ symmetry, decorates the resulting domain walls with an SPT in 1 dimension lower, and finally proliferates the domain walls to restore the symmetry $G$. One can also do this with defects of higher codimension. A mathematical counterpart of this general construction is the Smith homomorphism  discussed below.

The classification of free fermionic SPT phases also predicts a unique nontrivial $d=3$ SPT phase. It can be realized by a spin-polarized $p\pm ip$ superconductor \cite{Furusakietal,Kitaev}. It is characterized by the presence of a pair of counter-propagating massless Majorana fermions on the edge of the SPT phase.

In four space-time dimensions, the cobordism classification says that fermionic SPT phases are labeled by elements of $\ZZ_{16}$. Free fermionic SPTs in $d=4$  are classified by $\ZZ$ \cite{Furusakietal,Kitaev}, but with interactions turned on $\ZZ$ collapses to $\ZZ_{16}$ \cite{CFV}. The generator of $\Omega_4^{Pin^+}=\ZZ_{16}$ is the eta invariant of a Dirac operator \cite{Stolz}. The corresponding free fermionic SPT phase can be realized by a spin-triplet superconductor \cite{Furusakietal,Kitaev}. It is characterized by the property that on its boundary there is a single massless Majorana fermion. 

Two layers of the basic phase can be constructed from the $d=2$ phase with time-reversal symmetry $T^2=1$, via the map
$$
[\cap w_1^2]^:\Omega^{\rm \nu Pin^+}_4 \to \Omega^{\rm \nu Pin^-}_{2}.
$$
The map sends a the bordism class of a manifold $X$ on the left hand side to the bordism class of a codimension-2 submanifold of $X$ representing
$w_1^2(TX)$. From the physical viewpoint, the order 8 phase with $T^2=(-1)^F$ can be obtained from the trivial SPT phase by decorating certain codimension 2 defects (self-intersections of time-reversal domain walls, see the 3d case above) with the order 8 $D=1$ phase with $T^2=1$, i.e. the Kitaev chain.

Eight copies of this fermionic SPT phase are equivalent to a bosonic SPT phase with time-reversal symmetry and the effective action $\int w_1^4$ (the bosonic SPT phase predicted by group cohomology, see \cite{Kapcob1}). To show this, we need to show $8\eta = w_1^4$ for every $Pin^+$ 4-manifold. The space $\RR\mathbb{P}^4$ generates the $Pin^+$ bordism group in 4 dimensions, so every such manifold $X$ is $Pin^+$ bordant to a disjoint union of $k$ $\RR\mathbb{P}^4$s. Since $\eta$ is a $Pin^+$ bordism invariant, it follows $8\eta(X) = 8k\eta(\RR\mathbb{P}^4)$. Now $w_1^4$ is also a bordism invariant, so $w_1^4(X) = kw_1^4(\RR\mathbb{P}^4)$. Thus, we just need to show $8\eta(\RR\mathbb{P}^4) = w_1^4(\RR\mathbb{P}^4)$. We know the left hand side is $-1$ since the bordism group is $\ZZ/16$ and $\eta$ generates the dual group, and it is simple to show $w_1^4(\RR\mathbb{P}^4)=-1$ as well. The equivalence of these two phases was also argued in \cite{Wang14}.

Note that the eta-invariant cannot be written as an integral over a Lagrangian density $\mathcal{L}$ naturally associated to a lattice configuration on the underlying manifold $M$. In particular, if we have a covering map, we can pullback configurations to the cover. If the Lagrangian density were to simply pull back, then the action would just be multiplied by the number of sheets of the cover. However, for $M = \RR\mathbb{P}^4$ the eta-invariant associated to the standard Dirac operator is order 16 but trivial for its orientation double cover, $S^4$.

This signals that the effective field theory requires a certain amount of non-locality. It cannot have a description where each $Pin^+$ structure corresponds to a lattice configuration which respects covering maps of spacetimes up to gauge transformations.

It is interesting to note that the topological $Pin^+$ bordism group in 4d is $\ZZ_8$ rather than $\ZZ_{16}$. There is a manifold homeomorphic to the smooth generator $\RR \mathbb{P}^4$ but not smoothly $Pin^+$ cobordant to it which has a $\ZZ_{16}$ invariant equal to 9 as opposed to $\RR \mathbb{P}^4$'s 1 (these numbers are equal mod 8). The eta-invariant distinguishes these two manifolds. Since the classification of topological insulators in 3+1d is known to be at least $\ZZ_{16}$, this example shows that the spacetimes relevant to these systems always carry smooth structure. 

\subsection{$T^2=1$}

We propose that interacting fermionic SPT phases protected by time-reversal symmetry with $T^2 = 1$ are classified by the $Pin^-$ cobordism groups with $U(1)$ coefficients. In low dimensions the $Pin^-$ bordism groups are \cite{KirbyTaylor}
\begin{equation}
\Omega_{1}^{Pin^-}(pt) = \ZZ_2, \ \ \ \Omega_{2}^{Pin^-}(pt) = \mathbb{Z}_8, \ \ \ \Omega_{3}^{Pin^-}(pt) = 0, \ \ \ \Omega_{4}^{Pin^-}(pt) = 0, \nonumber \\ 
\end{equation}
and the cobordism groups are their Pontryagin duals.

In one space-time dimension, fermionic SPT phases are classified by $\ZZ_2$. This is easily interpreted in physical terms: the non-degenerate ground state can be either bosonic or fermionic, without breaking $T$.

In two space-time dimensions, a $Pin^-$ structure can be thought of as a $\ZZ_4$-valued quadratic enhancement of the intersection form which in the oriented (Spin) case is even and reduces to our description above\cite{KirbyTaylor}. Such a form $q$ satisfies $q(x+y) = q(x) + 2x\cap y + q(y) \mod 4$, where $2 x \cap y$ represents the mod 2 intersection of $x$ and $y$ mapped to $\ZZ_4$. The bordism group $\Omega^{Pin^-}_2=\ZZ_8$ is generated by $\RR\mathbb{P}^2$. The effective action is a generalization of the Arf invariant, the Arf-Brown-Kervaire invariant:
\begin{equation}
S(q) = \frac{1}{\sqrt{|H^1(X,\ZZ_2)|}} \sum_{A \in H^1(X,\ZZ_2)} \exp(2\pi i q(A)/4).
\end{equation}
It takes values in $\ZZ_8\in U(1)$.
If $q(x)$ is even for all $x$ (that is, if $q$ is $\ZZ_2$-valued), it reduces to the Arf invariant.
This situation occurs when the space-time is orientable.

From the physical viewpoint, the generator of $\ZZ_8$ is the Majorana chain, which can be regarded as a time-reversal invariant system with $T^2=1$. Time-reversal protects the dangling Majorana zero modes from being gapped out in pairs. 
Instead, interactions can only gap out octets, yielding a $\mathbb{Z}_8$ classification of phases labeled by the number of dangling modes \cite{Fid10}. Moreover, four copies of the Majorana chain with $T^2=1$ have states on the boundary on which T acts projectively, $T^2=-1$ \cite{Fid10}; hence, four copies of the basic fermionic SPT phases with time-reversal $T^2=1$ are equivalent to the basic bosonic SPT phase in $d=2$ with time-reversal symmetry. We can easily see this from the cobordism viewpoint. The generator of the $Pin^-$ bordism group in $d=2$ is $\RR\mathbb{P}^2$, so the fourth power of the generator of the cobordism group is $-1$ for this spacetime (here we are thinking about $\ZZ_8$ as a subgroup of $U(1)$). Meanwhile, $w_1^2$ is also $-1$ on $\RR\mathbb{P}^2$. Since both of these are $Pin^-$-bordism invariants, they are equal on all $d=2$  spacetimes.

As with the eta-invariant discussed above, the Arf-Brown-Kervaire invariant does not admit a local expression. There is a $\nu Pin^+$ structure on $\mathbb{R}\mathbb{P}^2$ for which the Arf-Brown-Kervaire invariant is a primitive 8th root of unity. However, the corresponding $Spin$ structure on the orientation double cover $S^2$ has Arf-Brown-Kervaire invariant 1 (the unique $Spin$ structure on the 2-sphere extends to a 3-ball).

\section{Fermionic SPT phases with a unitary $\ZZ_2$ symmetry}

Let $g$ denote the generator of a unitary $\ZZ_2$ symmetry. There are two possibilities: either $g^2=1$ or $g^2=(-1)^F$. In this section we discuss the former possibility only; the other one is discussed in the next section.

We propose that interacting fermionic SPT phases with unitary $\ZZ_2$ symmetry $g$, $g^2=1$, are classified by 
$$
\Omega^d_{Spin, tors}(B\ZZ_2)=\Hom(\Omega_d^{Spin,tors}(B\ZZ_2),U(1))
$$ 
The analogous group in the bosonic case is $\Omega^d_{SO,tors}(B\ZZ_2)$. In all dimensions there is an isomorphism called the Smith isomorphism
$$
\tilde\Omega_d^{Spin}(B\ZZ_2)\to \Omega_{d-1}^{Pin^-}(pt),
$$
where on the left hand side we use the tilde to denote reduced bordism: the kernel of the forgetful map to $\Omega_d^{Spin}(pt)$. The torsion part of reduced bordism is dual to SPT phases which can be made trivial after breaking the symmetry. Not all SPT phases are of this sort. One could imagine that after breaking the symmetry the system is reduced to some non-trivial SRE like the Kitaev chain. In general,
$$
\Omega_d^{Spin}(BG) = \tilde\Omega_d^{Spin}(BG)\oplus \Omega_d^{Spin}(pt),
$$
so these effects can be separated consistently and the Smith isomorphism is enough to classify the $G=\ZZ_2$ phases. This splitting fails if any elements of $G$ are orientation reversing or if $G$ acts projectively on fermions.

The Smith isomorphism is defined as follows. Starting with a Spin manifold $X$ and some $A \in H^1(X,\ZZ_2)$ representing a class on the left hand side, we produce a submanifold $Y$ Poincar\'e dual to $A$. (That we can do this is a special fact about codimension 1 classes with $\ZZ_2$ coefficients. Not all homology classes are represented by submanifolds.) The manifold $Y$ is not necessarily orientable. The Spin structure on $TX$ restricts to a Spin structure on $TY \oplus NY$, where $NY$ is the normal bundle of $Y$ in $X$. In fact, $NY$ is classified by the restriction of $A$ to $Y$. We compute
$$
0 = w_1(TX)|_Y = w_1(TY \oplus NY) = w_1(TY) + A,
$$
so on $Y$ the gauge field $A$ restricts to the orientation class, ie. the $\ZZ_2$ symmetry is orientation-reversing for $Y$. We also have
$$
w_2(TY \oplus NY) = w_2(TY) + w_1(TY)^2,
$$
so the Spin structure on $X$ becomes a $Pin^-$ structure on $Y$.

Physically, the submanifold $Y$ Poincar\'e dual to $A$ represents $\ZZ_2$ domain walls. The dual map from the $Pin^-$ cobordism of a point in $d-1$ dimensions to the $Spin$ cobordism of $B\ZZ_2$ in $d$ dimensions has the following physical meaning. Picking an element of the  $Pin^-$ cobordism group gives us a $d-1$-dimensional fermionic SPT with time-reversal symmetry $T^2=1$. To obtain a $d$-dimensional SPT, we decorate $\ZZ_2$ domain walls with this $d-1$-dimensional SPT and then proliferate the walls.

The inverse map can be described via compactification. One takes the $d$-dimensional SPT on a spacetime which is a circle bundle over the $d-1$-dimensional (perhaps unorientable) spacetime. This circle bundle is the unit circle bundle of the orientation line plus a trivial line, and is therefore oriented. We give the gauge field nontrivial holonomy around this circle and compactify. The effective field theory in $d-1$ dimensions is the $d-1$-dimensional SPT phase with time-reversal symmetry.

Fermionic SPT phases with a unitary $\ZZ_2$ symmetry have not been much studied in the physics literature. In one space-time dimension, they are classified by $\ZZ_2\times\ZZ_2$, since the ground state can be either bosonic or fermionic, as well as $g$-even or $g$-odd. In three space-time dimensions, Levin and Gu \cite{LevinGu} argued that fermionic SPT phases with $\ZZ_2$ symmetry and zero thermal Hall conductance are classified by $\ZZ_8$. Both of these results agree with the cobordism approach. 

\section{Fermionic SPT phases with a general symmetry}

A choice of spin structure gives a lift of the oriented frame bundle $P_{SO(d)}$ to a spin frame bundle $P_{\rm Spin(d)}$. Neutral Dirac spinors are sections of the bundle $S$ associated to this one by the complex spin representation. For Dirac spinors charged under some $G$ representation $\rho$, they are sections of the tensor bundle
$$
\psi \in \Gamma(S\otimes_\CC A^*\rho),
$$
where $A^*\rho$ denotes the vector bundle associated to the gauge bundle by $\rho$. Bosonic observables are composed of fermion bilinears which are sections of the tensor square of this bundle or the tensor product of this bundle with its dual. These are composed of integral spin representations of $SO(d)$ and exterior powers of $\rho^2$.

However, the situations where the spacetime is not a spin manifold are still physically important if $\rho$ is a projective representation. That is, while the spin frame bundle $P_{Spin(d)}$ or charge bundle $A^*\rho$ may not exist, the tensor product above does. For example, when $\rho$ is a half-charge representation of $G=U(1)$ the choice of a tensor product bundle is the same as a $Spin^c$ structure with determinant line $\rho^2$. One also knows that such a $Spin^c$ structure is the same as a spin structure on $TX \oplus A^*\rho^2$. 

One way to deal with this situation is to regard the fermions in $d$ dimensions as dimensional reduction of fermions in $d+n$ dimensions. Under such a reduction, the rotation group $SO(n+d)$ decomposes into $SO(d)\times SO(n)$ (for the moment we assume that the $d$-dimensional theory does not have orientation-reversing symmetries, and accordingly the $d$-dimensional space-time is orientable). We imagine that the symmetry group $G$ is embedded into $SO(d)$, and denote by $\xi$ the $G$-representation in which the $n$-vector of $SO(n)$ transforms. We can think of $\xi$ as a particular $G$-bundle over $BG$. Spinors in $d+n$ dimensions are elements of an irreducible module over the Clifford algebra built from $\RR^n\oplus \xi$. 

Consider now the theory on a curved space-time $X$ equipped with a $G$-bundle $A$. As usual, we can think of $A$ as a map from $X$ to $BG$, defined up to homotopy. To define the theory on such a space-time we must specify the bundle in which the fermions take value. This bundle must have the same rank as the spinor of $SO(d+n)$ and be a module over a bundle of Clifford algebras $T^*X\oplus A^*\xi$. Such a bundle is called a spin structure on the $SO(d+n)$-bundle $T^*X\oplus A^*\xi$. 

If some of the symmetries are orientation-reversing, we need to allow $X$ to be unorientable, so that the structure group of the tangent bundle is $O(d)$ rather than $SO(d)$. But we can compensate for this by embedding $G$ into $O(n)$ so that the generators of the Clifford algebra transform as a vector of $SO(d+n)$. Then fermions must take values in the irreducible Clifford module over the corresponding bundle of Clifford algebras, as before.

This discussion leads us to the following proposal Given a bosonic symmetry group $G$, and its representation $\xi$, fermionic SPT phases in $d$ space-time dimensions with this symmetry structure are classified by
$$
\Omega^d_{Spin}(\flat BG,\xi),
$$
a cobordism theory dual to the torsion part of the bordism theory of $d$-manifolds $X$ with a map $A:X\to \flat BG$ (the gauge field) and a spin structure on $TX\oplus A^*\xi$. It is important for continuous groups to use $\flat BG$ rather than $BG$ since gauging the $G$ symmetry means coupling to a flat $G$ gauge field. Turning on curvature for the gauge field requires a kinetic term which is non-canonical. One model for $\flat BG$ is to take the classifying space of $G$ as a discrete group. For finite $G$ this is of course automatic.

The data $(G,\xi)$ may seem to depend on some uphysical details, like the embedding of $G$ into $SO(n)$, but one can show that cobordism groups thus defined depend only on $w_1(\xi):G \to \ZZ_2$, which picks out the orientation reversing elements, and $w_2(\xi)\in H^2(G,\ZZ_2)$ \cite{Barrera-Yanez}, which determines how $G$ is extended by fermion parity.

Let us illustrate this with some examples. For $G=\ZZ_2$, first there is the trivial representation, for which this twisted cobordism group is the ordinary ones classifying fermionic SPTs with an internal $\ZZ_2$ symmetry acting honestly on the fermions, so the total symmetry group is $\ZZ_2 \times \ZZ_2^F$.

The other irreducible is the 1d sign representation. For this representation we have $w_1$ equal to the generator of $H^1(B\ZZ_2,\ZZ_2)$, this being the determinant of the representation, and $w_2 = 0$ since ths representation is 1 dimensional. We compute
$$
w_1(TX \oplus A^*\xi) = w_1(TX) + A^*w_1(\xi) = w_1(TX) + A,
$$
so an orientation of $TX \oplus A^*\xi$ identifies $A$ with the orientation class of $X$. We also have
$$
w_2(TX \oplus A^*\xi) = w_2(TX) + w_1(TX)A^*w_1(\xi) = w_2(TX) + w_1(TX)^2,
$$
a trivialization of which is a $Pin^-$ structure on $TX$. Thus,
$$
\Omega^d_{Spin}(B\ZZ_2, sign) = \Omega^d_{Pin^-}.
$$
Since $w_1(\xi) \neq 0$ and $w_2(\xi) = 0$ we interpret this group as classifying fermionic SPTs with an orientation-reversing symmetry such as time reversal which satisfies $T^2=1$. Note that the same group classifies SPT phases with a reflection symmetry squaring to 1.

We can also consider a sum of two sign representations, for which we have $w_1(\xi) = 0$ and $w_2(\xi) \neq 0$. This gives a bordism theory of oriented manifolds with $A^2 = w_2(TX)$. This symmetry structure is that associated to an orientation preserving symmetry such as particle-hole symmetry which squares to the fermion parity.

The sum of three sign representations has both $w_1(\xi)$ and $w_2(\xi)$ nonzero. The cohomology of $B\ZZ_2$ implies also $w_2(\xi) = w_1(\xi)^2$. With this we compute
$$
w_1(TX \oplus A^*\xi) = w_1(TX) + A
$$
and
$$
w_2(TX \oplus A^*\xi) = w_2(TX) + A^2 + A^2 = w_2(TX).
$$
The first implies that $A$ equals the orientation class of $X$. The second says that a spin structure on $TX\oplus A^*\xi$ is the same as a $Pin^+$ structure on $TX$. Thus
$$
\Omega^d_{Spin}(B\ZZ_2, 3\times sign) = \Omega^d_{Pin^+}.
$$
Therefore fermionic SPT phases with an orientation reversing $\ZZ_2$ symmetry squaring to the fermion parity are classified by $Pin^+$ cobordism.

For $G=U(1)$ there are no continuous representations with $w_1 \neq 0$ and $w_2 \neq 0$ for a continuous representation precisely when the sum of charges is odd. In this case $A^*w_2(\xi)$ is the mod 2 reduction of the gauge curvature $F_A$. A spin structure on $w_2(TX\oplus A^*\xi)$ is therefore the same thing as a $Spin^c$ structure with determinant line $F_A$. Note that these are not the $Spin^c$ cobordism groups studied in most of the mathematical literature since we require the determinant line to be flat.

For $G = U(1) \times \ZZ_2$ we now have representations where the $\ZZ_2$ is orientation reversing. For example, consider $\xi = charge\ 1 \otimes trivial \oplus trivial \otimes sign$. For this representation, $w_1(\xi)$ is the map to $\ZZ_2$ which is trivial on $U(1)$ and the identity on $\ZZ_2$. We also find
$$
w_2(TX\oplus A^*\xi) = w_2(TX) + w_1(TX)^2 + F_A.
$$
If we instead used three copies of the sign representation, we would have
$$
w_2(TX\oplus A^*\tilde\xi) = w_2(TX) + F_A.
$$
It may first appear that these give different cobordism theories, but note that $w_1(TX)^2$ lifts to an integral class, so a redefinition of the $U(1)$ field produces an equivalence between the two bordism groups. This is the same redefinition used in \cite{Wang14} to show that the $T^2 =1$ and $T^2 = (-1)^F$ classifications agree, a result verified here in cobordism. This is also reflected in the uniqueness of the $Pin^c(d)$ group and we find that both types of phase are classified by $Pin^c$ bordism with flat determinant line.

Now consider $G = U(1)\rtimes \ZZ_2$ with $\ZZ_2$ acting by conjugation. This group can be thought of as $SO(2)\rtimes \ZZ_2 = O(2)$. Consider first the standard 2d representation $\xi$. For this, $w_1(\xi)$ is the determinant $O(2) \to \ZZ_2$ and $w_2(\xi)$ is the obstruction to finding a section of
$$
Pin^+(2)\to O(2),
$$
ie. it is the class in group cohomology $H^2(BO(2),\ZZ_2)$ classifying $Pin^+(2)$. The ring $H^*(BO(2),\ZZ_2)$ is generated by the universal Stiefel-Whitney classes $w_1$ and $w_2$, and $w_2(\xi)$ is the universal $w_2$. This representation corresponds to $T^2 = 1$ since $T^2=1$ in $Pin^+(2)$.

One can also consider $T^2 = (-1)^F$ by using the representation $\tilde\xi = \xi + 2\times sign$. For this, $w_1(\tilde\xi)=w_1(\xi)$, but $w_2(\tilde\xi)$ is the universal $w_2 + w_1^2$, which differs from the other representation, demonstrating that these two classifications differ when time reversal does not commute with $U(1)$.

\section{Decorated Domain Walls}

The formulation above in terms of the global symmetry representation $\xi$ carried by fermion bilinears highlights some interesting features of the so-called decorated domain wall construction described in \cite{ChenLuVishwanath}. 

Let us start with a concrete example with a unitary $\ZZ_2$ symmetry which squares to fermion parity. We consider in 1+1d a massless Dirac fermion $\psi$ coupled to a massless real scalar $\phi$ by the Yukawa coupling $\phi\bar\psi\psi$. The $\ZZ_2$ symmetry we consider is $\phi \mapsto -\phi$, $\psi \mapsto \gamma^5 \psi$, where $\gamma^5 = i\gamma^0\gamma^1$. We condense $\phi$, making the domain wall infinitely heavy, and we consider the system on a line with boundary conditions $\phi \to \infty$ on the right and $\phi \to -\infty$ on the left. Then there is a domain wall at some fixed position and a $\psi$ zero mode bound to it. The point is to define the quantum mechanical theory of this zero mode, we need to pick a time direction. The ambient 2 dimensional space-time is oriented, so we can orient the domain wall if we can orient its normal direction. This orientation has to come from which side has the boundary condition $\phi \to \infty$ and which side has the boundary condition $\phi \to -\infty$. We choose some convention such as the $\phi \to \infty$ side is the positive side and thus orient the domain wall. However, if we now perform a global $\ZZ/2$ symmetry transformation, it swaps the boundary conditions but not the ambient orientation, so it reverses the time direction on the domain wall.

We can understand what happened in terms of the representation theory of $\ZZ_2$. We have to find the representation of $\ZZ_2$ on the fermion bilinears. There are three of them: $\bar\psi\psi$, $\bar\psi\gamma^\mu\psi$, and $\bar\psi\gamma^5\psi$. The first and the last transform as the sign representation, while the vector is invariant. Thus, $\xi$ is two copies of the sign representation. As calculated in the previous section, we have $w_1(\xi) = 0$, $w_2(\xi) \neq 0$, meaning that we have a unitary symmetry squaring to the fermion parity. Recall now that to define fermions in a background $G$ gauge field $A$ we used a spin structure on $TX \oplus A^*\xi$. If $Y$ is a curve in $X$, then $TX = TY \oplus NY$. If $Y$ is Poincar\'e dual to $A$, then $NY$ is $A^*{\rm sign}$. Altogether then, our fermions restricted to $Y$ are defined using a spin structure on $TY \oplus A^* \sign\oplus A^*\xi = TY \oplus A^*(\xi \oplus \sign)$. That is, for the fermions on the domain wall, $\xi$ is effectively shifted by a copy of the sign representation. To understand how the domain wall operators have different transformation properties, consider the operator $\bar\psi\gamma^x\psi$, where $\gamma^x$ is the Clifford operator in the oriented normal to the domain wall. Because we need to use the oriented normal to define this operator in the 0+1d theory, we have $\gamma^x \mapsto -\gamma^x$ under the $\ZZ_2$ symmetry, so $\bar\psi\gamma^x\psi \mapsto -\bar\psi\gamma^x\psi$, contributing another copy of the sign representation. So for the example just described we now have $\xi' = 3\times \sign$. Accordingly, as computed in the previous section, $w_1(\xi')\neq 0$ and $w_2(\xi')\neq 0$, so on the domain wall, $\ZZ_2$ has been transmuted into an orientation-reversing symmetry squaring to the fermion parity.

We pause before considering the general case to note that this feature is independent of dimension and for $\ZZ_2$ has an interesting order 4 periodicity as we cycle through each type of $\ZZ_2$ symmetry:
\[... \to {\rm unitary,\ squaring\ to\ }(-1)^F \to {\rm antiunitary, squaring\ to\ }(-1)^F \]
\[\to {\rm unitary,\ squaring\ to\ }1\to {\rm antiunitary,\ squaring\ to\ }1 \to ...\]
where the arrow denotes restriction to the $\ZZ_2$ domain wall.

Now let's consider the general case of $G$ symmetry with fermion bilinear representation $\xi$. In order to study a domain wall as we did above, we need a real scalar $\phi$ transforming in some 1 dimensional representation of $G$. This is the same as a group homomorphism $\sigma:G \to \ZZ_2$. In any decorated domain wall picture, the degrees of freedom bound to the wall are defined in a symmetry broken regime where the domain wall is infinitely tense. This choice of regime corresponds to the choice of $\sigma$. After the coupling is made in this regime, domain walls are again proliferated, restoring the $G$ symmetry. For such a $\sigma$, in the phase where $\phi$ is condensed, the domain wall $Y$ is Poincar\'e dual to the $\ZZ/2$ gauge field $\sigma(A)$ induced from the $G$ gauge field $A$. In particular, the normal bundle to the domain wall is $A^*\sigma$. Thus, the ambient spin structure restricts to a spin structure on $TY \oplus A^*(\xi \oplus \sigma)$. That is, the $G$ symmetry properties on the domain wall correspond to the fermion bilinear representation $\xi \oplus \sigma$.

In terms of cobordism groups, every map $\sigma:G \to \ZZ_2$ induces a map
\[ \Omega_d^{\rm Spin}(BG,\xi) \to \Omega_{d-1}^{\rm Spin}(BG,\xi\oplus\sigma)\]
and thus a map
\[ \Omega^{d-1}_{\rm Spin}(BG,\xi\oplus\sigma) \to \Omega^d_{\rm Spin}(BG,\xi).\]
Note that domain walls may be coupled to different degrees of freedom in different symmetry breaking sectors, corresponding to adding the images of maps from different $\sigma$s.

It is also possible to couple domain defects of higher codimension through higher dimensional representations $\sigma$. These representations may be irreducible over $\RR$, so this procedure is not always equivalent to merely iterating the above construction. For example, if $G = \ZZ_4$, then we can take $\sigma$ to be the 2-dimensional representation rotating the plane by $\pi/2$. This representation is irreducible over $\RR$ since the eigenvectors of this rotation are imaginary. This representation defines a map
\[ \Omega^{d-2}_{\rm Spin}(B\ZZ_4,\xi \oplus \sigma) \to \Omega^{d}_{\rm Spin}(B\ZZ_4,\xi).\]
One must be careful in defining these maps in general, however, since not every homology class Poincar\'e dual to $A^*\sigma$ is representable by a manifold if the dimension of $\sigma$ is too large. Happily this does not occur until the ambient dimension is at least 6.

\section{Concluding remarks}

We have seen that cobordism correctly predicts the known classification of interacting fermionic SPT phases in $D\leq 3$ with $\ZZ_2$ symmetry, either unitary or anti-unitary. We find that for $0\leq D\leq 3$, all phases are realized by free fermions. However, in higher  dimensions new phenomena occur. First of all, while the classification of free fermionic SPT phases with a fixed symmetry exhibits mod 8 periodicity in dimension \cite{Kitaev}, in the interacting case there is no periodicity. Second, the deviations from the free fermionic classification occur for high enough $D$, but the precise point depends on the symmetry group. For example, for SPT phases with time-reversal symmetry $T$, $T^2=(-1)^F,$ deviations start at $D=3$. For SPT phases with no symmetry beyond $(-1)^F$ deviations start at $D=6$. (In $D=6$ the free fermionic classification predicts $\ZZ$, but in the interacting case it is $\ZZ\times\ZZ$ because there are two different gravitational Chern-Simons terms possible based on the Pontryagin numbers $p_1^2$ and $p_2$, respectively.) 

Third, while in low dimensions the effect of interactions is to truncate the free fermionic classification, in high enough dimension inherently interacting fermionic SPT phases appear. For example, in $D=7$ free fermionic SPT phases with time-reversal symmetry $T$, $T^2=(-1)^F$, are classified by $\ZZ$, while the cobordism approach predicts $\ZZ_2\times\ZZ_{32}$. The latter group is not a quotient of the former, so truncation alone cannot explain the discrepancy. The most likely interpretation is that $\ZZ_{32}$ is a truncation of $\ZZ$, while the $\ZZ_2$ factor corresponds to an inherently interacting fermionic SPT phase. Similarly, in $D=6$ there should exist inherently interacting fermionic SPT phases with only fermion parity as a symmetry. 

We have found that the correct classification requires the use of smooth manifolds rather than topological manifolds. It would be interesting to determine whether there is some physical difference between the smooth and piecewise linear categories.

We find also that the fermionic SPT effective action has a degree of non-locality that was not present in the case of bosonic SPTs. For $D=1$, the effective action can be written in terms of a sum over an auxiliary $\ZZ_2$ gauge field. It is tempting to interpret it as a gauge field which couples to the fermion parity, but this needs to be tested. We leave this and the determination of possible boundary behaviors of fermionic SPT phases to further work.

\section*{Appendix}

In this appendix we discuss the quantization of the coefficient of the gravitational Chern-Simons action. For all topological facts used here, the reader may consult \cite{MilnorStasheff}. Let $X$ be an oriented 3-manifold whose tangent bundle is equipped with a connection $\omega$. We can take $\omega$ to be a Levi-Civita connection for some Riemannian metric on $X$, so $\omega$ can be thought of as an $SO(3)$ connection. 

We define the gravitational Chern-Simons action to be
$$
S_{grav}(\omega)=\frac{\kappa}{192\pi}\int_M\Tr (\omega d\omega +\frac{2}{3}\omega^3).
$$
The choice of the normalization coefficient will be explained shortly. This formula is only schematic, since $\omega$ is not a globally-defined 1-form, in general. A more precise definition requires choosing a compact oriented 4-manifold $M$ whose boundary is $X$ (this is always possible, since $\Omega_3^{SO}(pt)=0$). We also extend $\omega$ to $X$ and define
$$
S_{grav}^X(\omega)=\frac{k}{192\pi}\int_X \Tr R\wedge R.
$$
We need to ensure that $\exp(i S_{grav}^X(\omega))$ does not depend on the choice of $X$ or the way $\omega$ is extended from $M$ to $X$. If we choose another $X'$ with the same boundary $M$, the difference between the two ways of defining the gravitational Chern-Simons action is
$$
\frac{k}{192\pi}\int_{X'\cup\bar X} \Tr R(\omega)\wedge R(\omega),
$$
where $\bar X$ is $X$ with orientation reversed, and $R(\omega)$ is the curvature 2-form of $\omega$. This expression can be rewritten as
\begin{equation}\label{sgravdifference}
\frac{\pi k}{24} p_1(X'\cup \bar X)=\frac{\pi k}{8}\sigma(X'\cup \bar X).
\end{equation}
Here $p_1(Y)$ denotes the first Pontryagin number of a closed oriented 4-manifold $Y$, $\sigma(Y)$ denotes its signature, and we used the Hirzebruch signature theorem $p_1(Y)=3\sigma(Y).$ Since the signature is an integer,  we conclude that $\exp(iS_{grav}(\omega))$ is well-defined provided $k$ is an integer multiple of $16$. This determines the quantization of the thermal Hall conductivity for $d=3$ bosonic SPTs with time-reversal symmetry.

Now suppose $M$ is given a spin structure. We can exploit it to define $\exp(i S_{grav})$ for arbitrary integral $k$. We merely require the spin structure to extend to $X$. It is always possible to find such an $X$, since $\Omega_3^{Spin}(pt)=0$. The difference between $S_{grav}^X(\omega)$ and $S_{grav}^{X'}(\omega)$ is again given by (\ref{sgravdifference}). Since now $X'\cup \bar X$ is a closed spin 4-manifold, we can appeal to the Rohlin theorem which says that the signature of a closed spin 4-manifold is divisible by 16, and conclude that $\exp(iS_{grav}(\omega))$ is well-defined if $k$ is integral. This determines the quantization of the thermal Hall conductivity for $d=3$ fermionic SPTs with time-reversal symmetry. Note that in the fermionic case the quantum of conductivity is 16 times smaller than in the bosonic case.

\end{document}